\pdfoutput=1
\documentclass{JINST}
\usepackage{graphicx}
\usepackage{enumitem}
\usepackage{mathtools}  
\usepackage{caption}
\usepackage{subcaption}

\usepackage{subcaption}
\setlist[itemize]{noitemsep, topsep=0pt}

\title{GRAPH - An readout ASIC for large MCP based detectors}

\author{A.~Seljak$^a$ $^e$ $^f$\thanks{Corresponding author.}, J.~Vallerga$^b$,  G. ~Liu$^d$ $^f$, R.~Raffanti$^c$, G. S.~Varner$^f$\thanks, \\
	\llap{$^a$}Jozef Stefan Institute, \newline  Jamova 39, 1000 Ljubljana, Slovenia\\
	\llap{$^b$}Space Sciences Laboratory,University of California,\newline 7 Gauss Way, Berkeley, CA 94720, USA \\
	\llap{$^c$}Techne Instruments,\newline 4920 Telegraph Ave, Unit. G, Oakland, CA 94609, USA\\
	\llap{$^d$}SLAC National Accelerator Laboratory,\newline 2575 Sand Hill Rd, Mailstop 0094, Menlo Park,  California  94025, USA \\
	\llap{$^e$}Former - Fondazione Bruno Kessler - FBK, \newline  Via Sommarive 18, 38123 Povo, Italy\\
	\llap{$^f$}Former - University of Hawai'i at Manoa, Department of Physics and Astronomy,\newline 2505 Correa rd., Honolulu, HI  96822, USA\\
	\llap{} \thanks~ Deceased

	E-mail: \email{Andrej.Seljak@ijs.si}}

\abstract{

	We present a programmable 16 channel, mixed signal, low power readout ASIC, having  the project historically named Gigasample Recorder of Analog waveforms from a PHotodetector (GRAPH). It is designed to read large aperture single photon imaging detectors using micro channel plates for charge multiplication, and measuring the detector's response on crossed strips anodes to extrapolate the incoming photon position. 
	\\Each channel consists of a fast, low power and low noise charge sensitive amplifier, which provides a myriad of coarse and fine programmable options for gain and shaping settings.  Further, the amplified signal is recorded using, to our knowledge novel, the Hybrid Universal sampLing Architecture (HULA), a mixed signal double buffer memory, that enables concurrent waveform recording, and selected event digitized data extraction. The sampling frequency is freely adjustable between few~kHz up to 125~MHz, while the chip's internal digital memory holds a history 2048 samples for each channel,  with a  digital headroom of 12 bits.  
	\\An optimized region of interest sample-read algorithm allows   to extract the information just around the event pulse peak, while selecting the next event, thus substantially reducing the operational dead time. 
	The chip is designed in 130~$n$m TSMC CMOS technology, and its power consumption is around 47~$m$W per channel.

}


\begin{document}

	This paper is dedicated to the  memory of Prof. Dr. Gary S. Varner. He will be missed, his legacy cherished.

	\section{Introduction}
	
\begin{figure}[h!]
		\begin{subfigure}{0.5\textwidth}
			\centering
			\includegraphics[width=0.7\linewidth]{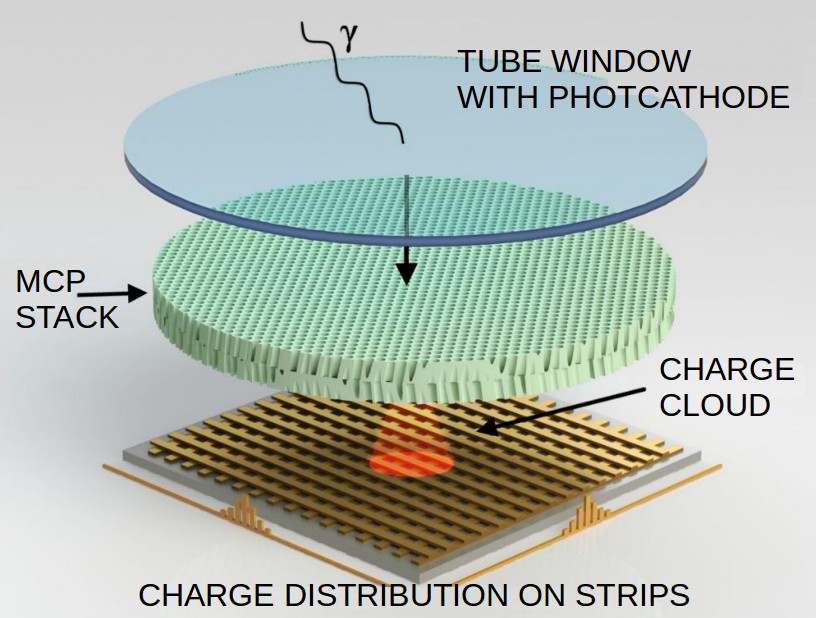}
			\caption{}
			\label{fig:1}
		\end{subfigure}%
		\begin{subfigure}{0.47\textwidth}
			\centering
			\includegraphics[width=1.15\linewidth]{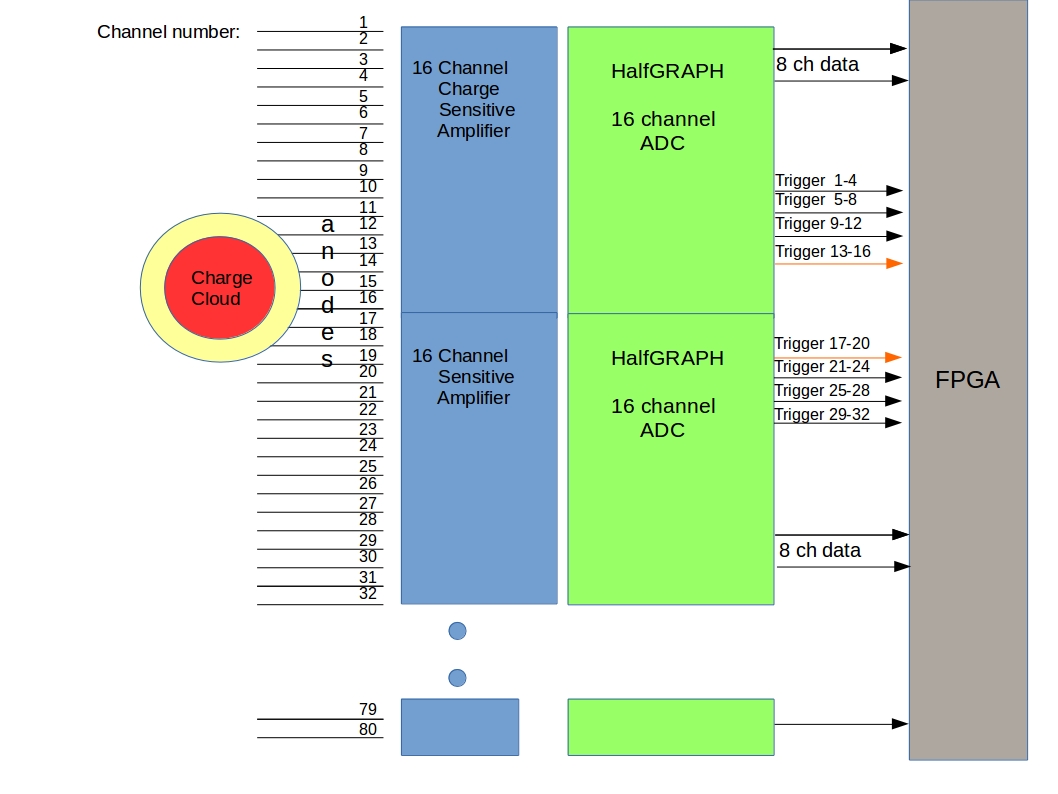}
			\caption{}
			\label{fig:2}
		\end{subfigure}
		\caption{(a) Schematic view of a detector using an  MCP stack and XS anode readout. (b) Scalable readout system concept using event localized triggering ability [\cite{HG2} ,\cite{System}.] }
		\label{fig:detector}
	\end{figure}

	High resolution single photon imaging detectors aimed at operation in low photon flux, are sought after for future space missions such as HabEX, CETUS and LUOVIR to name a few examples, and more generally regarded for Ring Imaging Cherenkov Detectors for  high energy physics experiments. 
	The here presented example of a single photon sensitive detectors uses a photo-cathode for quantum conversion into photo-electrons.  In a sealed (tubed) detector version\cite{Seigmund}, the photo-cathode would be usually deposited on the quartz entrance window of the enclosure, while for detectors that operate in free space, where vacuum is not an issue, and the entrance window is not needed, the Micro Channel Plate (MCP) itself can be coated with a photo-cathode of choice (Fig.~\ref{fig:detector}\subref{fig:1}). Relying on chevron stacked MCPs for signal amplification to detectable levels,  provides for long exposures having very low background noise on the images, compared to solid state counterparts. Furthermore, considering the recent  MCP technological advancements, very large size and long lasting imaging detectors can be obtained. 
	
	In principle, when a high voltage is applied to the MCP, it  creates an electric field that accelerates the quantum converted primary photo-electron along the micro pore, and at each collision on the walls, the multiplication effect takes place. The amount of charge exiting the MCP, namely the signal, is related to the number of collisions induced by the primary photo electron inside the stacked MCP plate. Since the multiplication avalanche doesn't  always begin at the same spot inside the micro pore, for a single photon the amount of charge may vary notably. Generally one can expect charge clouds containing 160~$femto$~Coulombs of charge, or the equivalent of about 1M electrons~(e$^-$) per detected photon.
	
	The charge cloud exiting the MCP is absorbed by an anode plane composed of an orthogonal array of 25~$\mu$m strips  (PXS), having a pitch of 625~$\mu$m  (Fig.~\ref{fig:detector}\subref{fig:1}), each being connected to a readout channel. By measuring the lateral charge distribution, the impinging photon entrance position can be calculated from the center of mass of the charge cloud \cite{John}, \cite{John2}. The main desired detector specifications for future UV astronomy applications are to have a large aperture (10x10cm), have the pixel resolution of 20~$\mu$m, being radiation hard, and most importantly, operate at very low power.

	Previous cross strip MCP detectors that achieved sub 20~$u$m~FWHM spatial resolution had electronics that were bulky and power hungry with dedicated fast ADCs per channel. High power and bulky electronics are exactly what one wants to avoid in space based UV detectors where mass and power are a limited resource.
	
	Our endeavor, to rise the Technology Readiness Level (TRL) of the detector's readout system from 4 to 6, lead us to initially design amplifiers and analog to digital converters separately using two different technologies. These chips were the CSAv3 (130~$n$m) \cite{CSAv3}, a 16 channel programmable amplifier, and the HalfGRAPH (250~$n$m)\cite{HG2}, a 16 channel giga sample per second waveform sampling and digitization ASIC. Both chips were used to construct a low power detector readout system \cite{System}  (Fig.~\ref{fig:detector}\subref{fig:2}),  which confirmed the proof of concept, reaching pixel resolution on the order of 30 $\mu$m and power consumption of 0.5W/cm$^2$, or dozen watts reading an area of 25 cm$^2$ \cite{large_det}

	Considering scaling the detector to very large formats, the PXS approach works favorably  in terms of power consumption\cite{large_det}. By duplicating the amount of readout channels, hence duplicating the power consumption, the sensitive area quadruples.  While the CSAv3 complied with the requirements, the HalfGRAPH architecture could be further optimized. Therefore, we decided to integrate the 16 channel CSAv3 amplifiers along with a novel dedicated ADC structure into a single GRAPH ASIC in 130 $n$m  \cite{Graph}.  
	
	In view toward space grade environmental specifications, aside from radiation hardness which is mission-driven, and not specifically addressed here,  the most concerning  is the power consumption. Heat is very difficult to dissipate, and the system's signal-to-noise ratio and sensitivity must be able to bring the desired pixel resolution toward 20~$u$m~FWHM at low MCP bias voltages.  Our previous work brought us to the conclusion, that the following parameters should be met in view of designing a new ASIC, and is summarized in table~\ref{table1}.

	\begin {table} [h!]
	\begin{center}
		\label{table1}
		\begin{tabular}{| l | l |}  
			\hline
			Input channels  &   16        \\ \hline
			
			Input dynamic range   & 0 - 50 fC  (300ke$^-$)             \\ \hline
			Target detector capacitance & 5 pF                   \\  \hline
			Amplifier gain                 & > 10 mV/fC  \\ \hline
			Shaping time         & rise/fall time 30~ns, pulse contained within 100~ns           \\ \hline
			ENC   &  $<=$ 1000 e$^-$ at 5 pF detector capacitance  \\ \hline
			Sampling frequency &  100~MHz \\ \hline
			ADC resolution    & 12-bit, system S/N in the order of 50 dB \\ \hline
			Event readout rate    &  1 M event/s/ASIC  (1 event = 6 samples*8 channels)    \\ \hline
			Power consumption    &  < 50 mW / channel        \\ \hline
		\end{tabular}
		\caption {General overview of system requirements }
	\end{center}
	\end {table}
	
	Space missions differ and  the technology selection doesn't offer a  generic solution, trade-offs  are made to fit the mission's purpose. In this respect, we note that there are at least two groups working on MCP based high resolution single photon imaging detectors. An implementation using a pixelated pattern, opposed to PXS, was under study using the Timepix  \cite{timepix},  with an  updated version with the Timepix4 ASIC \cite{Fiorini}, \cite{Ballabriga}. Another group uses the Beetle ASIC \cite{Diebold} to accomplish a similar endeavor using the PXS approach. 
	
	\section{GRAPH ASIC}

	\subsection{Architecture overview}

	Redesigning into an integrated system with amplifiers and ADCs  required moving to a smaller node (250 $n$m to 130 $n$m), which comes with some technical constrains. A  waveform sampling array is made of a large number of switches and  capacitors. These begin to channel leak charge from the storage capacitors back to the input, hence loosing the correct sampled value. This effect was found in simulation to be in the range of a few mV/us. Despite relying on short access and conversion times, opting for a random digitization moment of the event in the sampling array would have non-deterministic effects, hence produce large conversion errors.  In order to be able to  retain a high S/N ratio of the  system, the time between sampling and conversion should remain constant for all chips running synchronously. We explore, to our knowledge, a somewhat novel ADC architectural concept using  a mixed signal memory.  We named it the Hybrid Universal sampLing Architecture (HULA). 
	
	GRAPH ASIC \cite{Graph} is a mixed signal 16 channel  input, 12 bit digital data output system on chip designed for reading MCP detectors with crossed strip anodes.  For operation it requires two 1.2 V power supply rails, a 2.5 V for the LVDS transceiver operation, while the analog and digital grounds are split. It is developed in TSMC 130~$n$m technology on a die the size of roughly 8~by~9 mm and uses 128 interconnection pads. 
	
\begin{figure}[h!]
		\begin{subfigure}{0.5\textwidth}
			\centering
			\includegraphics[width=1\linewidth]{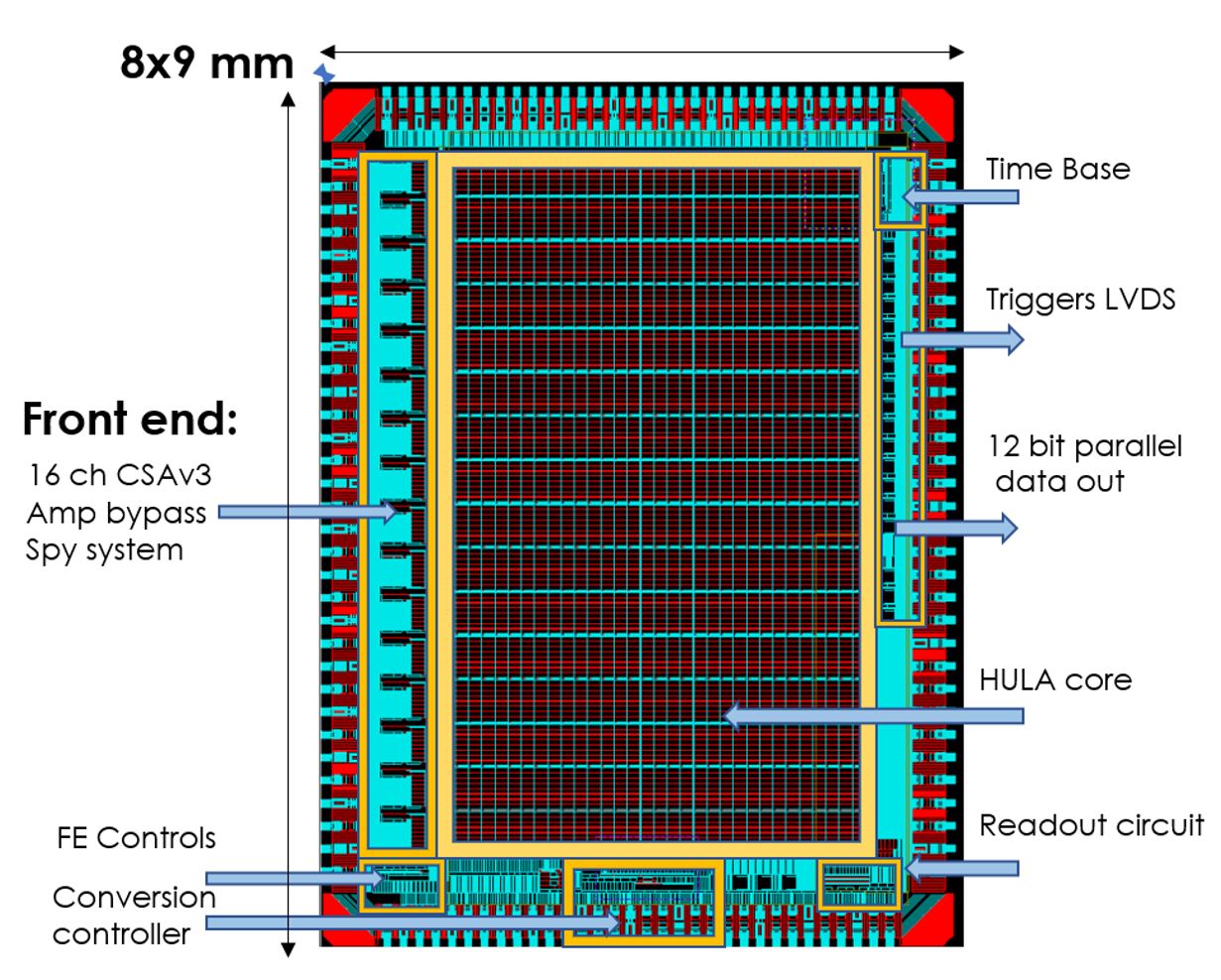}
			\caption{Spatial location of various electronic circuits}
			\label{fig:block_diagram}
		\end{subfigure}%
		\begin{subfigure}{0.47\textwidth}
			\centering
			\includegraphics[width=0.55\linewidth]{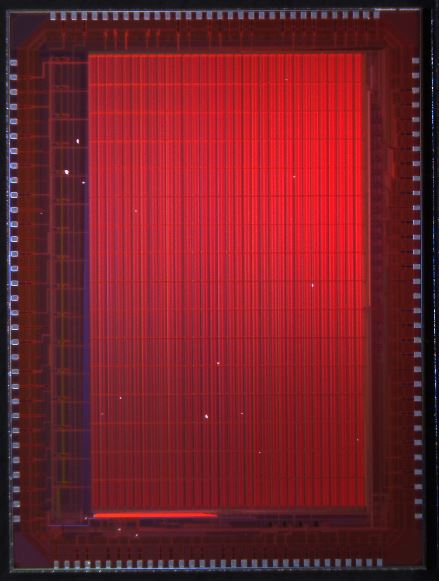}
			\caption{Die image}
			\label{fig:3}
		\end{subfigure}
		\caption{(a)ASIC layout   (b) GRAPH ASIC image }
		\label{fig:photo}
	\end{figure}

	The front end is composed of programmable CSAv3  \cite{CSAv3} amplifiers with an additional circuit that allows bypassing the CSA, if the ASIC is used in  applications,  where the amplified signal is  fed into the HULA ADC.  In order to control the sampling and conversion, the Timebase circuit generates the sampling vector address from a single input clock pin. It also generates internal synchronization signals that govern the conversion. In addition, an independent read system mechanism is available, which minimizes the readout dead time for this particular case of use.   The slow control system is composed of a  4 wire  SPI like receiver, accepring 1.2 V LVCMOS signals. Its command decoder  accepts frames being 64 bit long, containing the targeting internal address space and values to control circuits for various settings and internal biasing voltages (DACs).  Figure ~\ref{fig:block_diagram} depicts the internal structure of the ASIC and Fig~\ref{fig:photo}b the pictogram of the chip.

	With the GRAPH ASIC redesign  we address also some desired  improvements on the system level \cite {System} , and these are: 
	
	\begin{itemize}
		\item Reduce complexity and make it fully programmable (low pin count, low grade FPGA  required)
		\item Improve the event readout rate 
		\item Minimize the system's footprint to the point it could be scaled abutting smaller detectors
	\end{itemize}

	\subsection{Sampling windows and memory array}

\begin{figure}[h!]
		\begin{subfigure}{0.5\textwidth}
			\centering
			\includegraphics[width=1\linewidth]{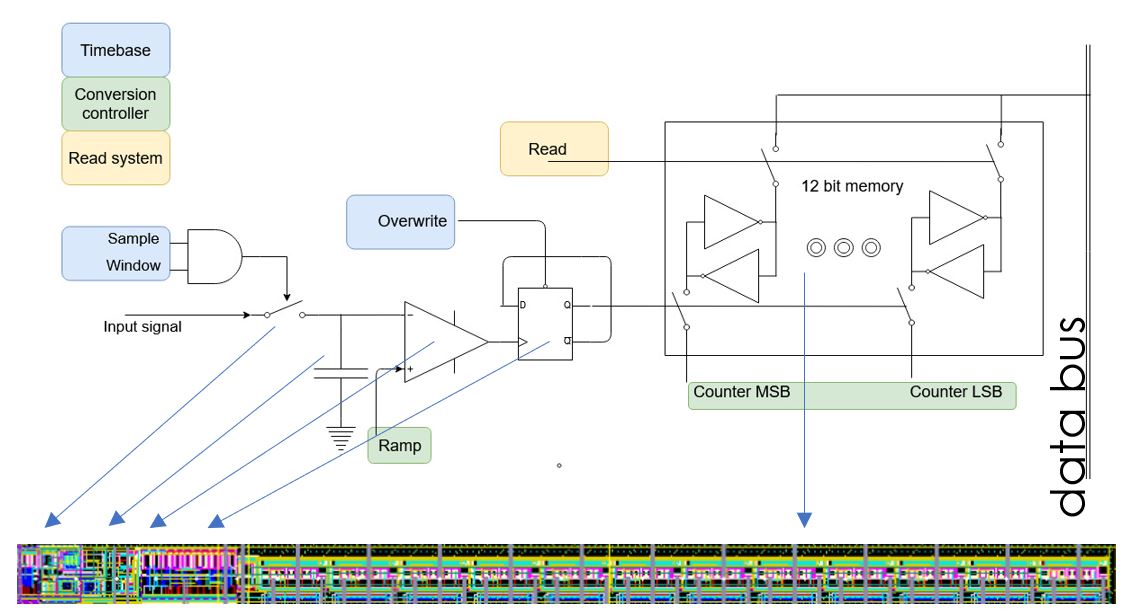}
			\caption{}
			\label{fig:block_diagram2}
		\end{subfigure}%
		\begin{subfigure}{0.47\textwidth}
			\centering
			\includegraphics[width=0.9\linewidth]{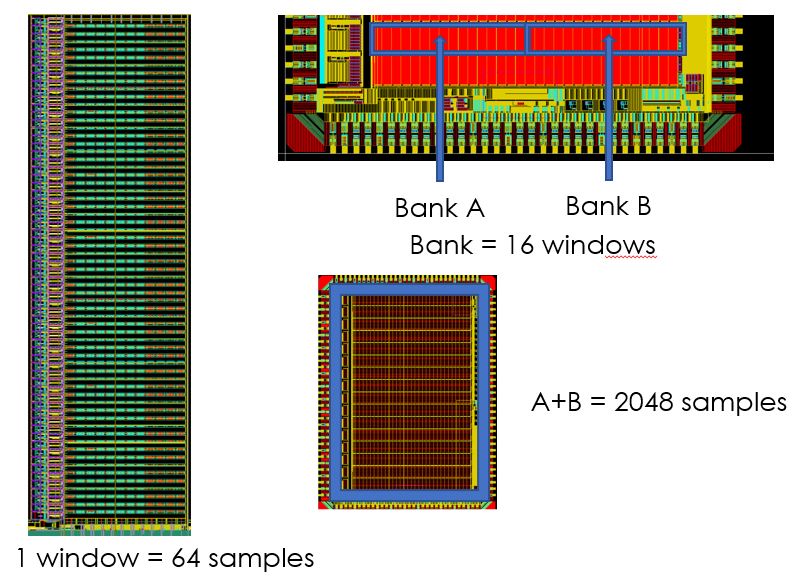}
			\caption{}
			\label{fig:3}
		\end{subfigure}
		\caption{(a)One sample of a mixed mode memory   (b) Window, Bank, Channel geometrical~placement }
		\label{fig:window}
	\end{figure}

	A single sample of the HULA architecture is built using a mixed signal memory, employing a small 30 fC capacitor for analog sampling, a comparator, an overwrite protection flip flop, and a 12 bit digital memory SRAM alike, shown in figure ~\ref{fig:block_diagram2}. Once the sample acquires its analog value, a conversion using the Wilkinson ADC approach will take place at some deterministic point in time. While in conversion, the 12 SRAM bits are connected to the bus and follow the counter value. As soon as the ramp value is above the signal in the capacitor, a D flip flop is set, which disconnects the 12 bit memory from the bus, retaining the last counter value. A separate bus is used to independently read the memory array.
	
	A single sample is stacked into a so-called vertically oriented window containing 64 samples, and there are two banks having 16 windows for each channel. Therefore the memory length of 2048 samples is available for every channel.  The memory banks are stacked horizontally, while channels are stacked vertically shown in \ref{fig:window}b.

	\subsection{Timebase generation circuit}
	
\begin{figure}[h]
		\centerline{
			\includegraphics[width=0.5\textwidth]{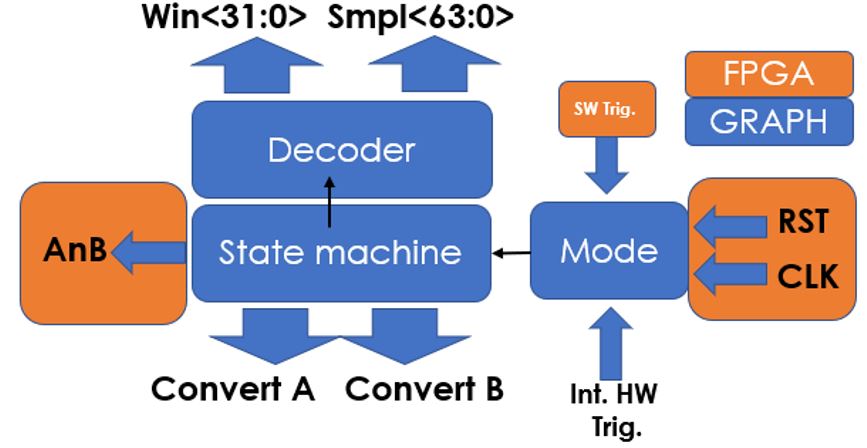}}
		\caption{Functional block diagram of the Timebase circuit.}
		\label{fig:timebase1}
	\end{figure}

\begin{figure}[h]
		\centerline{
			\includegraphics[width=1\textwidth]{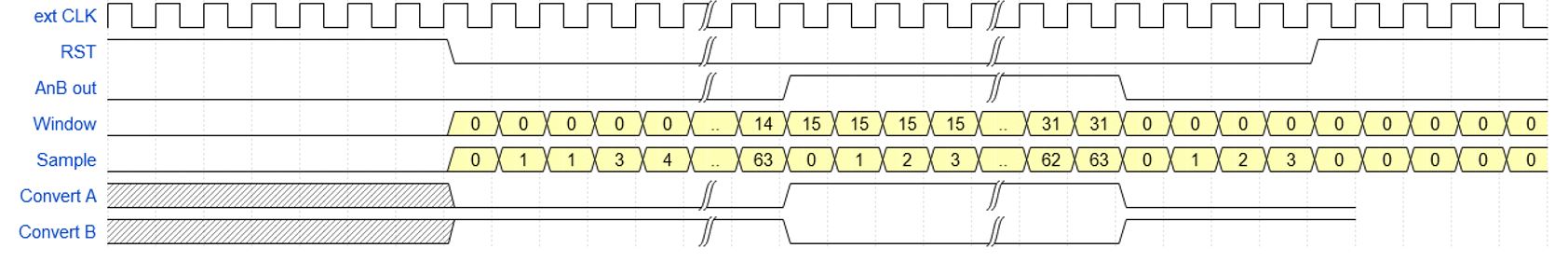}}
		\caption{Sampling signal sequence diagram.}
		\label{fig:timebase2}
	\end{figure}

	The Timebase generation circuit, figure \ref{fig:timebase1} is composed of a self looped serial shift register  being 64 bits in length, inside which a sample actuation signal of programmable width is let to circulate in a loop. When the external nCLR signal is released, the shift of this signal occurs driving the sampling action through the samples in the window at an externally provided clock.
	At each turnaround, a 5 bit counter increments by one, and its value represents the window which is addressed   for sampling, figure ~\ref{fig:timebase2}. 
	The signal from the shift register is directly fanned out to all windows arrays, while the decoded value in the 5 bit counter selects the window undergoing sampling ~\ref{fig:block_diagram2}. 
	In addition, the Timebase extrapolates a few more signals necessary for internal synchronization. An AnB signal is sent off the chip, when the transition from one bank to another happens and internal signals are set to control the conversion controller.

	\subsection{Conversion controller }
	
\begin{figure}[h!]
		\begin{subfigure}{0.5\textwidth}
			\centering
			\includegraphics[width=0.91\linewidth]{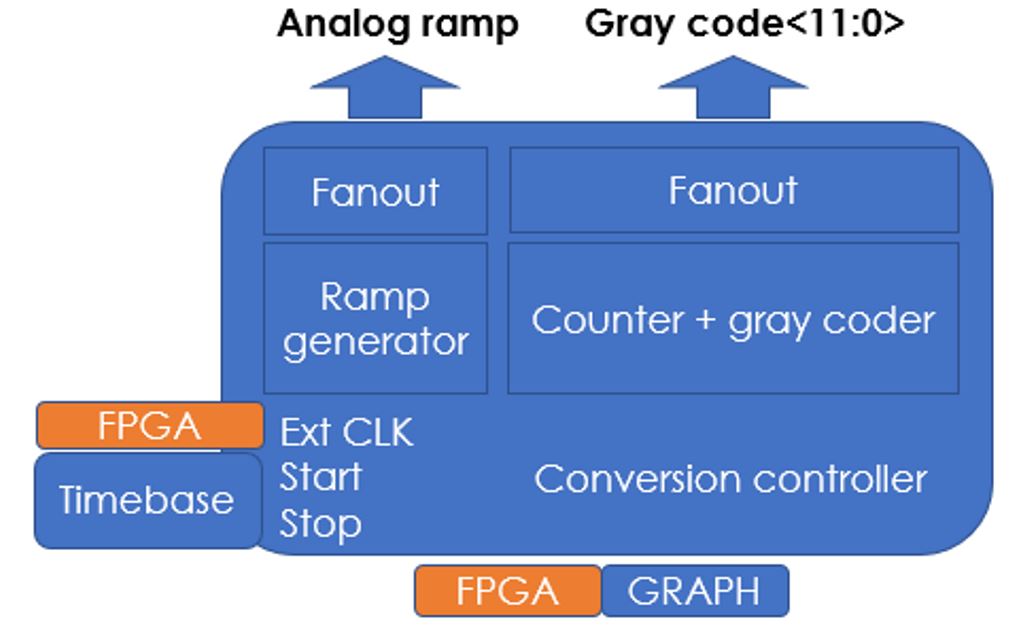}
			\caption{}
			\label{fig:conversion1}
		\end{subfigure}%
		\begin{subfigure}{0.4\textwidth}
			\centering
			\includegraphics[width=1\linewidth]{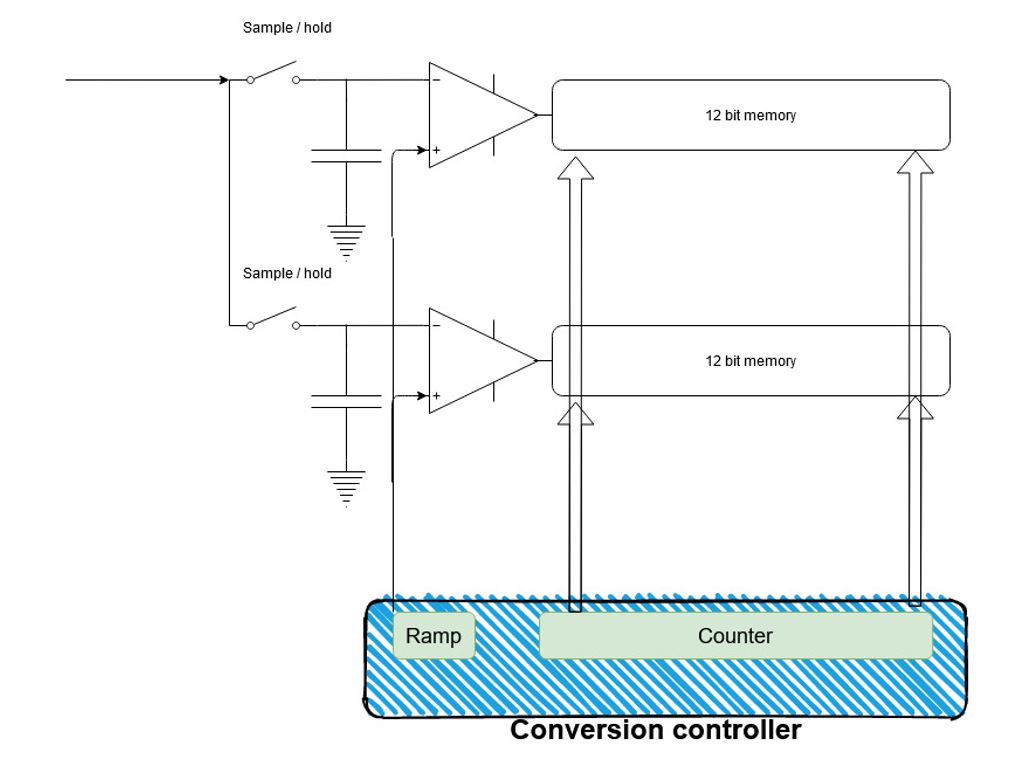}
			\caption{}
			\label{fig:3}
		\end{subfigure}
		\caption{(a)Functional block diagram of the Conversion controller (b) Conversion controller Bank fan out }
		\label{fig:conversion2}
	\end{figure}

	The main function of the conversion controller is to execute the conversion at a deterministic moment. It is composed of ramp generator, which relies on a programmable current source driven into a capacitor to create a voltage ramp, figure ~\ref{fig:conversion1}. A gray coded counter value is derived from a synchronous counter that revolves at the externally provided frequency clock. The 12 bit Gray code and analog ramp signal are further buffered for fanout over the 16 channels of the entire bank ~\ref{fig:conversion2}. Hence the ASIC has  two such controllers, one for each bank.  
	
	Once the Timebase unit signals the conversion request, the conversion controller will sync the external clock with the start of the ramp, and initiate the counter.  Conversely, when the conversion ends, the system self resets the counter, the ramp value and unlocks the memory allowing to be overwritten in the next iteration.

	\subsection{HULA operation}
	
\begin{figure}[h!]
		\centerline{
			\includegraphics[width=0.7\textwidth]{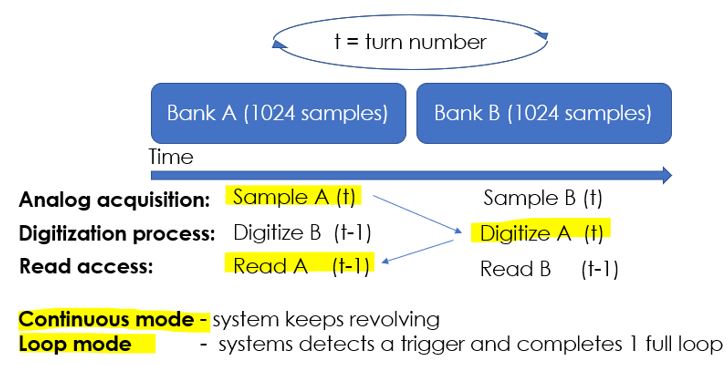}}
		\caption{HULA architecure opeartion }
		\label{fig:hula_diagram}
	\end{figure}

	HULA operates in a continuously revolving pattern. Once the sampling starts, the analog signal is sampled in the first Bank. As soon as the first Bank A has completed sampling, the AnB signal changes, signaling the transition to the external world, and automatically starts to sample bank B and puts the entire Bank A into conversion. When the Bank B has completed sampling, Bank A contains digitized data from the previous revolution which are available for the duration of new sampling period of Bank A, figure  ~\ref{fig:hula_diagram} . 
	
	Considering there are 1024 samples in each Bank, selecting a sampling clock of 125 MHz will require 8,192 $\mu$s to complete the sampling of one Bank. To reach 12 bit resolution in this time, the external clock should run at least at 2 $n$s period, hence at 500 MHz, and the ramp should be adjusted to cover the operating headroom. This allows to set an arbitrary sampling frequency, if a lower sampling rate is more adequate for a larger CSA shaping time. It also enables reading a much larger portion of the memory in the given conversion  time slot, covering more events.
	
	Two additional modes of operation are possible. One is the continuous mode, where the chip keeps overwriting data. And loop mode, when it  detects a trigger, it completes one revolution, and at this point, the entire memory is available for reading.

	\subsection{Data readout}

\begin{figure}[h!]
		\begin{subfigure}{0.5\textwidth}
			\centering
			\includegraphics[width=0.8\linewidth]{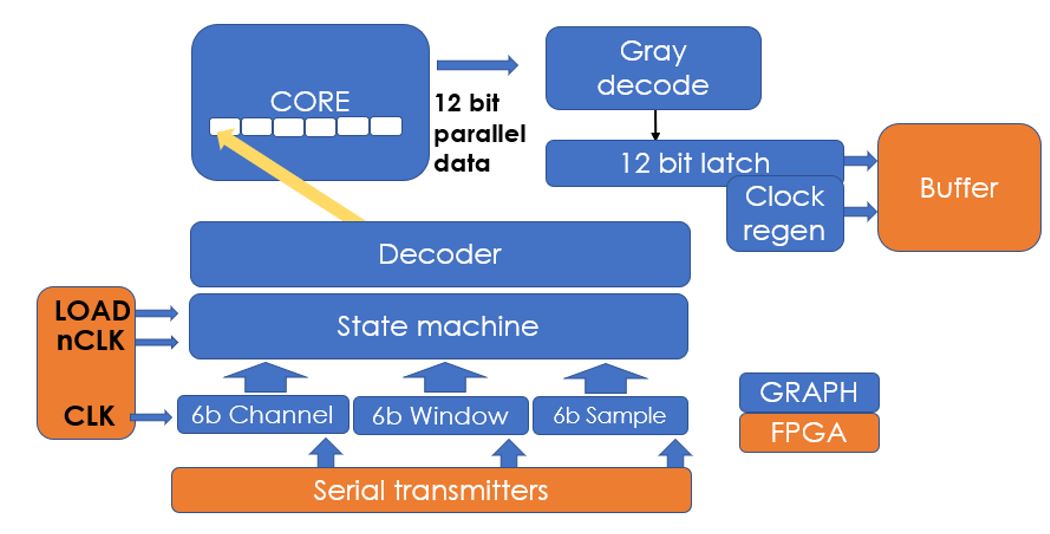}
			\caption{}
			\label{fig:1}
		\end{subfigure}%
		\begin{subfigure}{0.5\textwidth}
			\centering
			\includegraphics[width=1.15\linewidth]{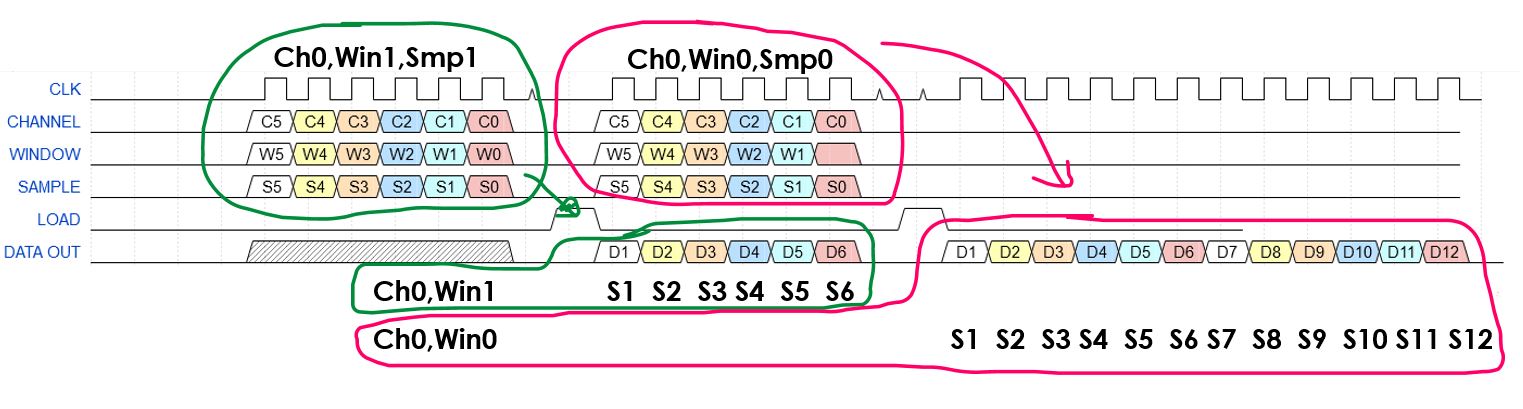}
			\caption{}
			\label{fig:2}
		\end{subfigure}
		\caption{(a)Functional block diagram of the data readout mechanism (b) Communication sequence diagram}
		\label{fig:readout}
	\end{figure}

	The mechanism for memory data readout is completely independent from sampling and conversion, and allows some flexibility in accessing values from the desired address. It is possible to read  consecutive samples  from the chosen address, while loading a new one, reducing dead time. This is achieved having a receiver  composed of 3 serial inputs (channel,window, sample), a corresponding read clock and load confirmation. When an address is shifted in the system and loaded, the value is an offset added  to a counter, which is driven by the read clock. At the rising edge of a clock cycle a new vector is decoded pointing to a single cell. This gets attached to the read bus and streamed to a 12 bit parallel output of the chip, which an FPGA latches at the opposite clock edge, figure ~\ref{fig:readout}a. Introducing a value of zero to all three registers, and continuing clocking, one would read sample 0 from channel 0, till the last sample of the channel 0. By continuing asserting the read clock,  sample 0 of channel 1 would follow and so on. In our application, we load the address of interest, and make the next 6 clock cycles to load new address, while collecting requested data from the previous call. Such access is envisioned for pulse shape extraction. The maximum transmission frequency tested so far was 60 MHz, which would drain the entire memory in loop mode in about 0.5 $m$s, collecting some 50kB of data.

	\newpage
	
	\subsection{Trigger section}
	
\begin{figure}[h!]
		\begin{subfigure}{0.55\textwidth}
			\centering
			\includegraphics[width=0.9\linewidth]{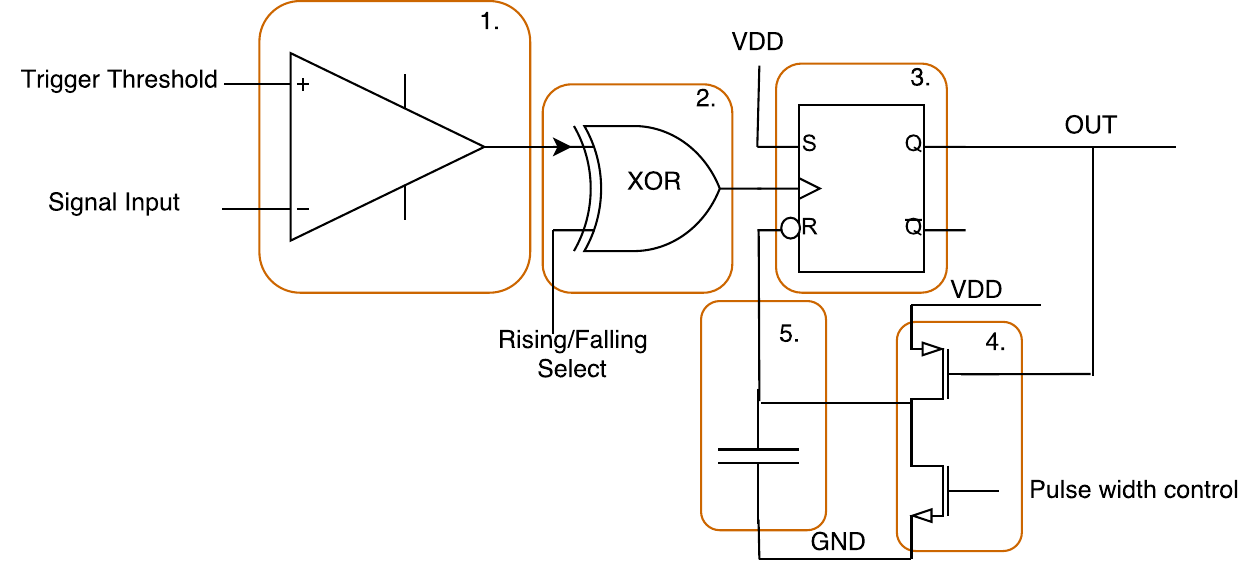}
			\caption{}
			\label{fig:trg0}
		\end{subfigure}%
		\begin{subfigure}{0.5\textwidth}
			\centering
			\includegraphics[width=0.8\linewidth]{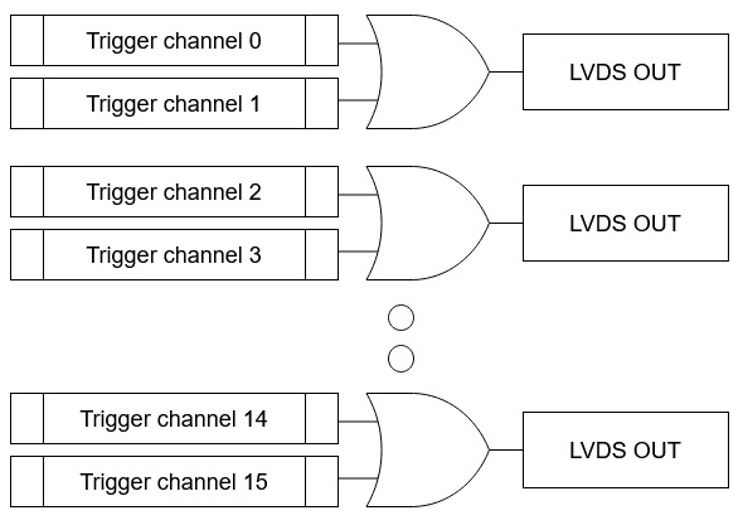}
			\caption{}
			\label{fig:3}
		\end{subfigure}
		\caption{(a) Channel trigger circuit  (b) Trigger outputs arrangement }
		\label{fig:trg1}
	\end{figure}

	Unlike operating in loop mode, in continuous mode triggers are key for the selection of events to pull from GRAPH memory. Each channel has a trigger circuit composed of a programmable threshold, a logical selection for firing on rising or falling edge of the input signal, and a self reset   logical signal (the actual trigger output), which pulse width can be set between a ns to few hundred ns by an internal DAC, figure~\ref{fig:trg0}. Adjacent triggers are OR-ed and fanned out toward the FPGA via an LVDS output \ref{fig:trg1}b.  To a degree it allows to discriminate the pulse hit, by setting a common threshold, while setting the width differently. This configuration simplifies  the required firmware for channel spatial charge distribution search. Ideally, the FPGA is only required to have an 11 bit counter (2048 sample space) being released by the nCLR signal as for all chips, and increments synchronously with the sampling clock. At trigger arrival the value is latched, and the readout processing sequence can estimate the events of interest to extrapolate the charge cloud signal form the MCP/anode. An additional possibility to enhance the functionality is by adding a TDC on the muxed triggers, and locate with better timming the absolute time of the photon arrival with sub $n$s time of arrival precision should be possible. 
	An additional option, the software trigger, is also available, and can be initiated via the slow control serial register. Both trigger options would stop the ASIC in loop mode operation, as well as, both can be masked to be overruled.

	\section{Results}
	
\begin{figure}[h!]
		\centerline{
			\includegraphics[width=0.5\textwidth]{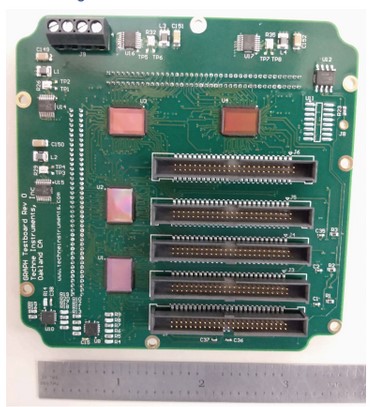}}
		\caption{Detector test board }
		\label{fig:board}
	\end{figure}

	For testing the ASICs, a PCB board that can be attached to a detector was produced, figure \ref{fig:board}. This  solution has currently 4 ASICs (2 per axis) wire bonded directly onto the PCB along with a minimal set of low drop regulators and a subsets of dual inline connectors to connect to an external FPGA board. A perpendicular connector made of two dual inline pin  headers is used to couple the readout system to the detector.   For testing, the sampling clock was set to 125MHz, and the Wilkinson conversion clock was set to 500 MHz.

\begin{figure}[h!]
		\begin{subfigure}{0.5\textwidth}
			\centering
			\includegraphics[width=0.91\linewidth]{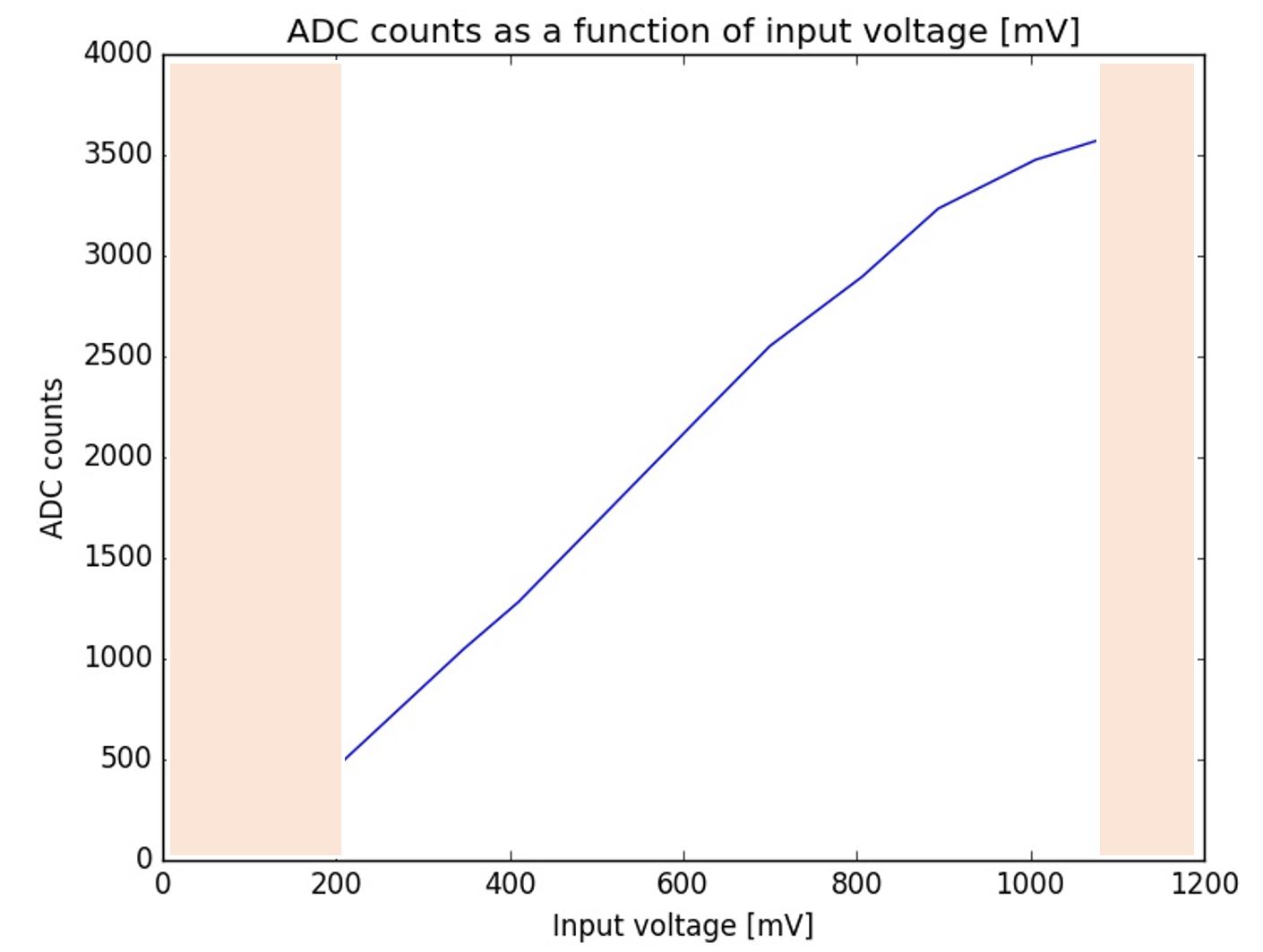}
			\caption{}
			\label{fig:1}
		\end{subfigure}%
		\begin{subfigure}{0.5\textwidth}
			\centering
			\includegraphics[width=0.91\linewidth]{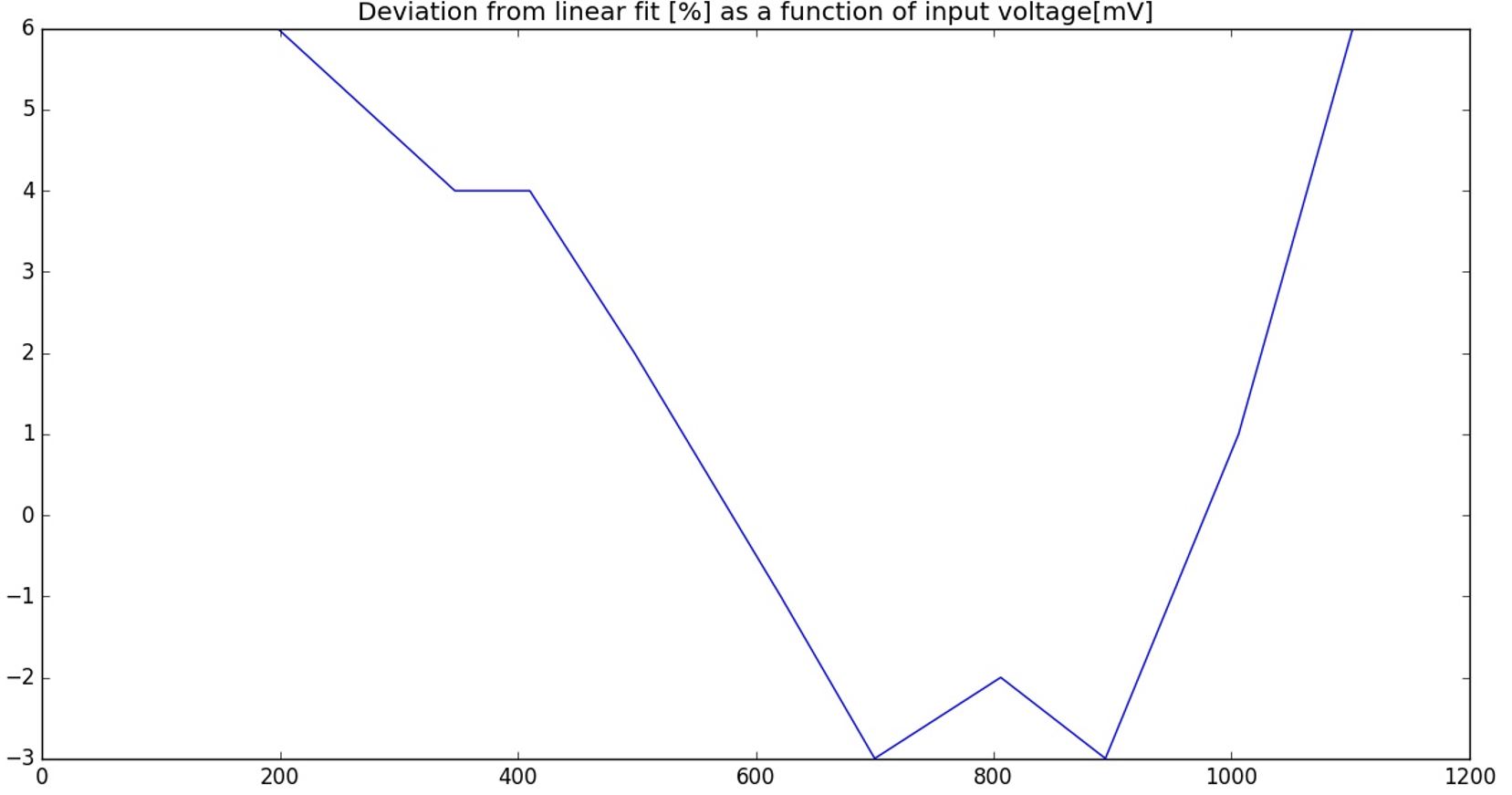}
			\caption{}
			\label{fig:2}
		\end{subfigure}
		\caption{(a)  GRAPH ADC counts as a function of input voltage. (b) Measured linearity error.}
		\label{fig:linearity}
	\end{figure}

	Initially, the CSA amplifier was skipped, and only the performance of the HULA core was examined  by carefully  tuning  the conversion slope to match the pedestal offset on both Banks.   Because the comparators in each memory cell cannot operate near the voltage supply rails, using a DC swap the usable  ADC headroom was extracted.  Although each ASIC is slightly different,  we found all of them operate safely within a 800 mV headroom covered by some 3000 ADC counts. The INL has found to be of few percent, figures~\ref{fig:linearity}a,\ref{fig:linearity}b.    
	
\begin{figure}[h!]
		\begin{subfigure}{0.5\textwidth}
			\centering
			\includegraphics[width=0.99\linewidth]{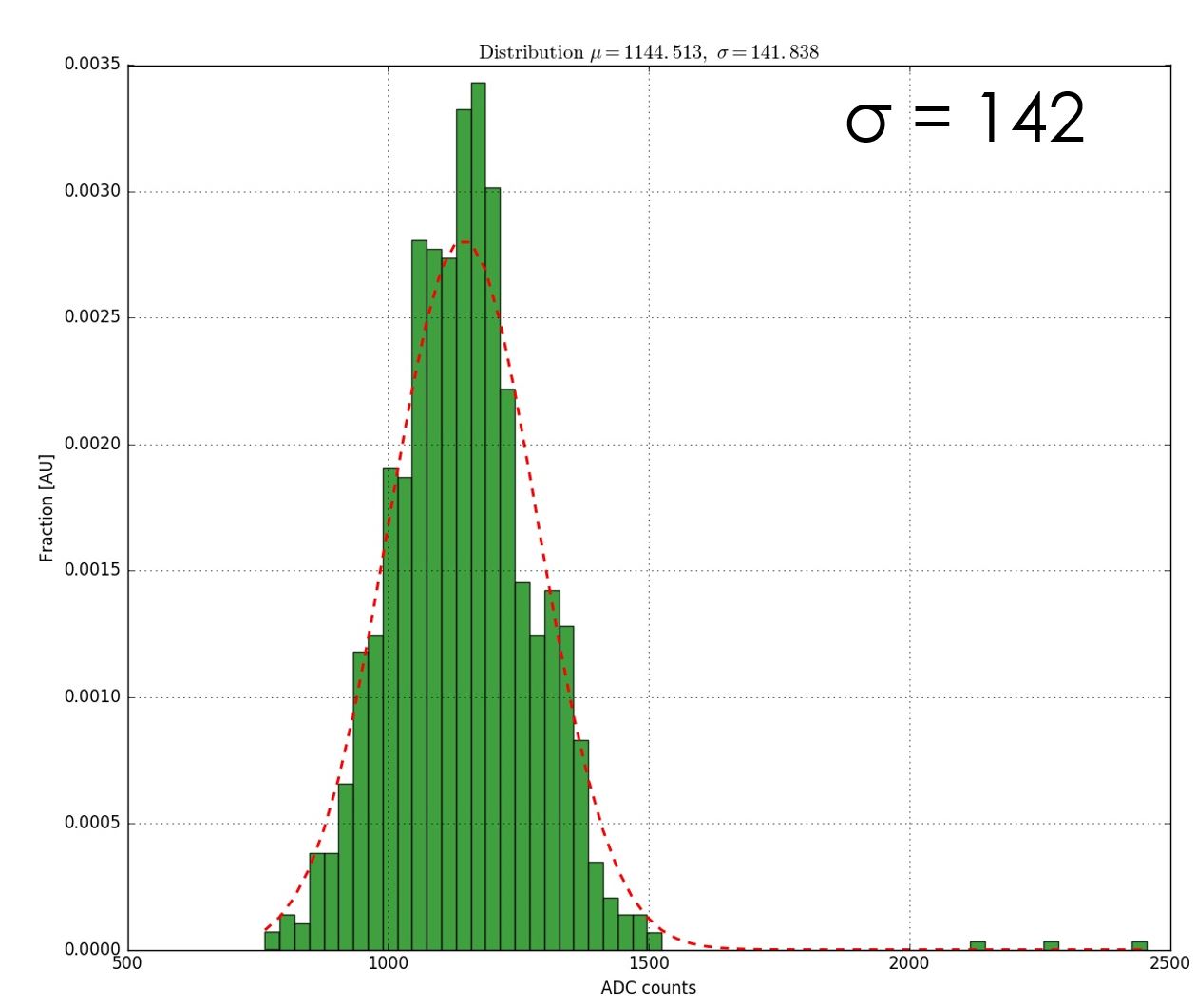}
			\caption{}
			\label{fig:1}
		\end{subfigure}%
		\begin{subfigure}{0.5\textwidth}
			\centering
			\includegraphics[width=0.91\linewidth]{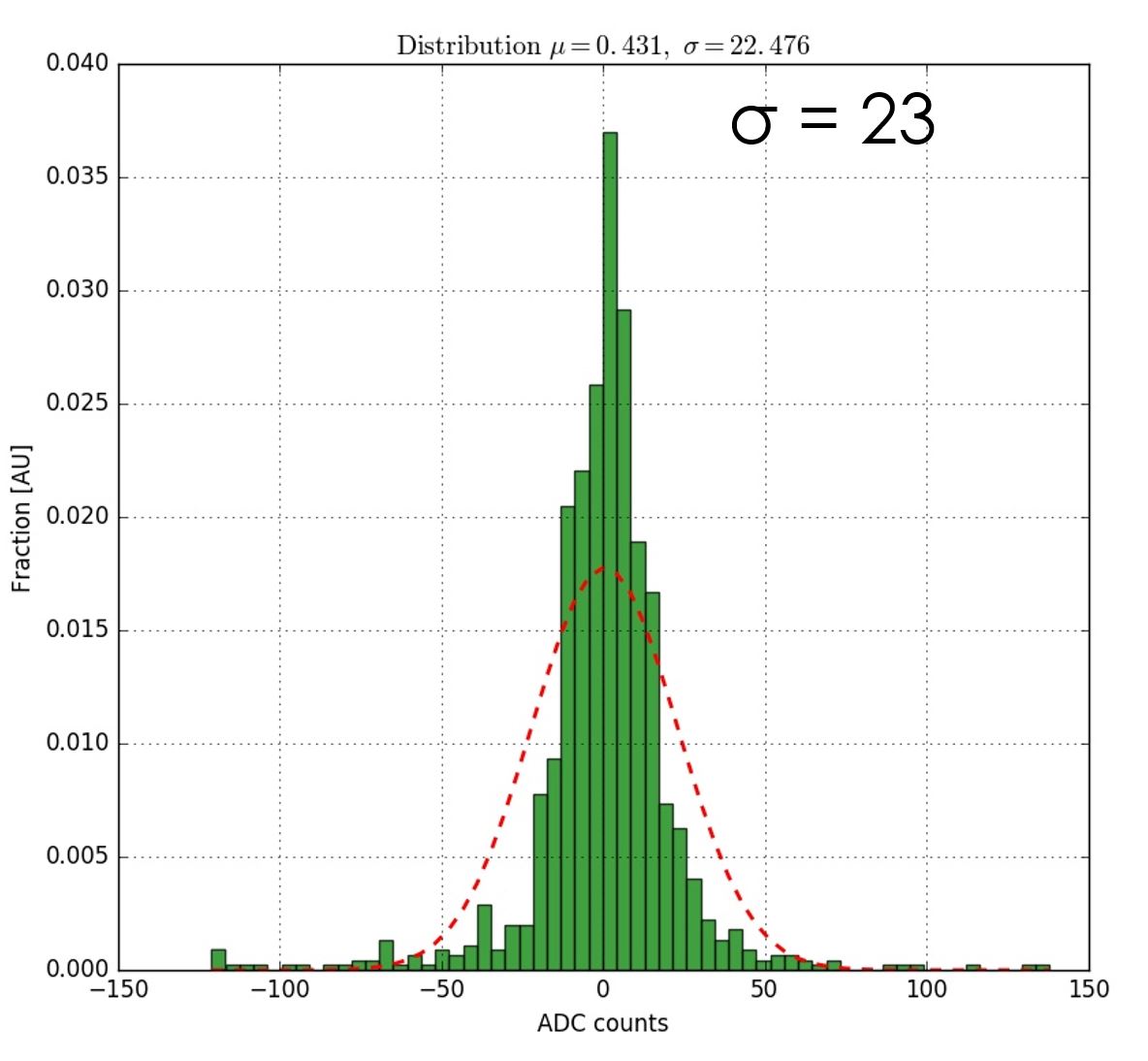}
			\caption{}
			\label{fig:2}
		\end{subfigure}
		\caption{(a)HULA noise contribution  (b) Pedestal corrected HULA noise contribution}
		\label{fig:linearity2}
	\end{figure}

	By applying a DC voltage of about 600 mV to the input, the recorded pedestal spread was found to be considerable, with a sigma of 142 counts, shown in figure \ref{fig:linearity2}a This somewhat expected result comes from a combination of factors. These are the systematic production mismatch of the capacitors  used in the sampling cell, the intrinsic firing point error of the comparator, and the location of cell itself inside the large memory array, as the ramp propagates through the fanout. A solid calibration method is to measure each cell about a 100 times and average its value. As a new measurement is taken, the averaged offset value is subtracted from the measurement. After doing so, the pedestal value is removed, the intrinsic offset of the cell is corrected reducing the spread to a sigma of 23 counts, figure~\ref{fig:linearity2}b. The ADC has an SNR of 20log*(3000/23) of around 42dB, with the sensitivity of about 6 mV. Although the system is set to use 12 bits of data, the ENOB figure is more around 7. The numbers obtained from the fit are spoiled due to large tails (figure~\ref{fig:linearity2}b) produced by a small fraction of large values outliers. In time domain these represent single samples that register a large value, and doesn't fire a trigger, so it seems are more related errors in conversion.  The effect isn't statically related to particular cells but rather happens sporadically.  We are sill investigating the cause.  By neglecting these single sample off errors, the sigma reduces toward 10 counts.

\begin{figure}[h!]
		\begin{subfigure}{0.5\textwidth}
			\centering
			\includegraphics[width=0.91\linewidth]{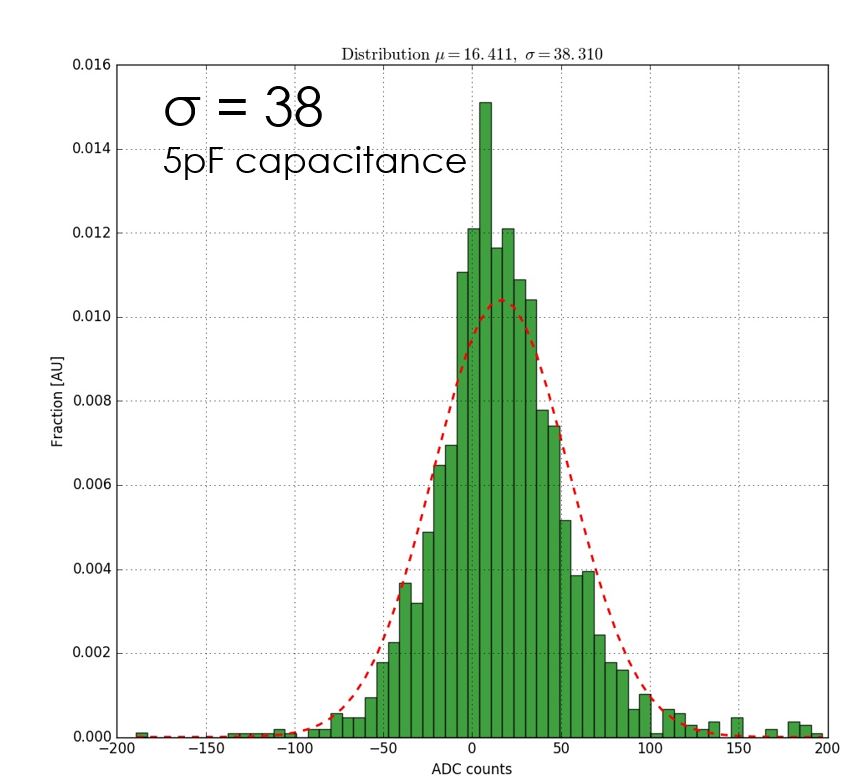}
			\caption{}
			\label{fig:1}
		\end{subfigure}%
		\begin{subfigure}{0.5\textwidth}
			\centering
			\includegraphics[width=1\linewidth]{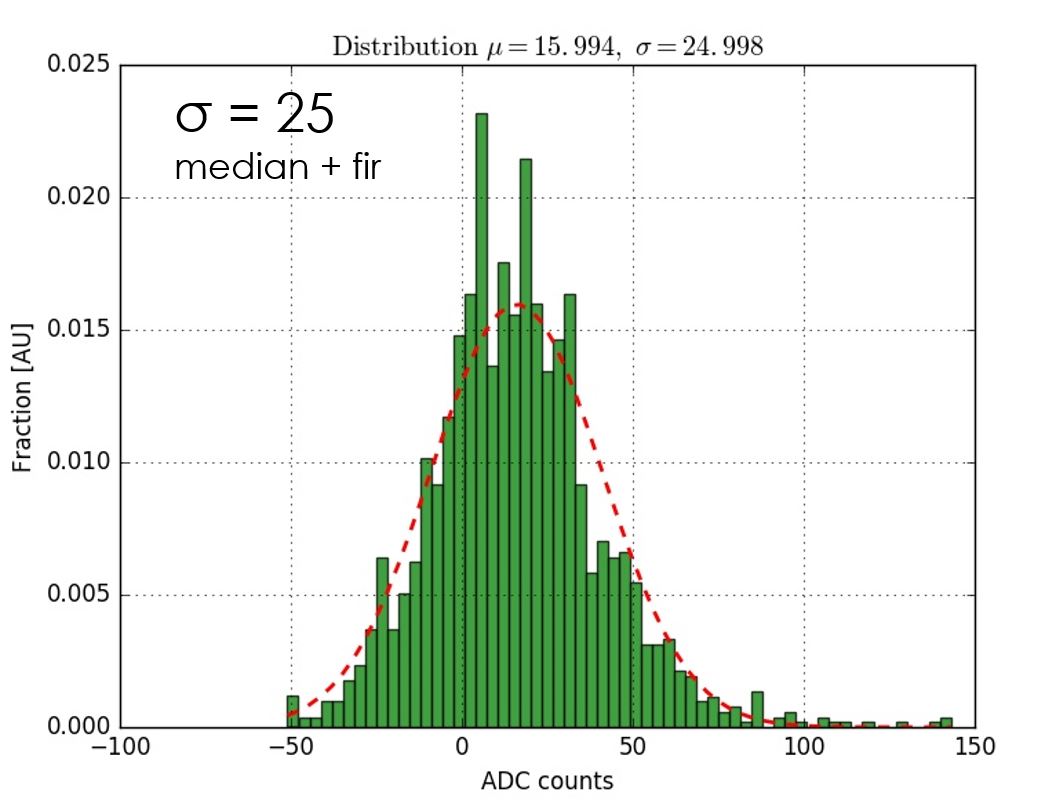}
			\caption{}
			\label{fig:2}
		\end{subfigure}
		\caption{(a)GRAPH ASIC noise contribution attached to the detector (b) Median and fir filter corrected result}
		\label{fig:linearity3}
	\end{figure}

	The system was connected to the detector, and the CSAv3 amplifiers were turned on having the calibration procedure being repeated. The amplifier noise generated by the 5 $p$F input capacitance has worsen the result to 38 ADC counts, figure~\ref{fig:linearity3}a. It was found that the noise contribution could be reduced by off line post processing the data using a combination of a median and FIR filter to some degree of success without degrading the signal shape,  figure~\ref{fig:linearity3}b.  None the less, the over all obtained raw results are close to the specifications and expected performance.

\begin{figure}[h!]
		\begin{subfigure}{0.5\textwidth}
			\centering
			\includegraphics[width=0.8\linewidth]{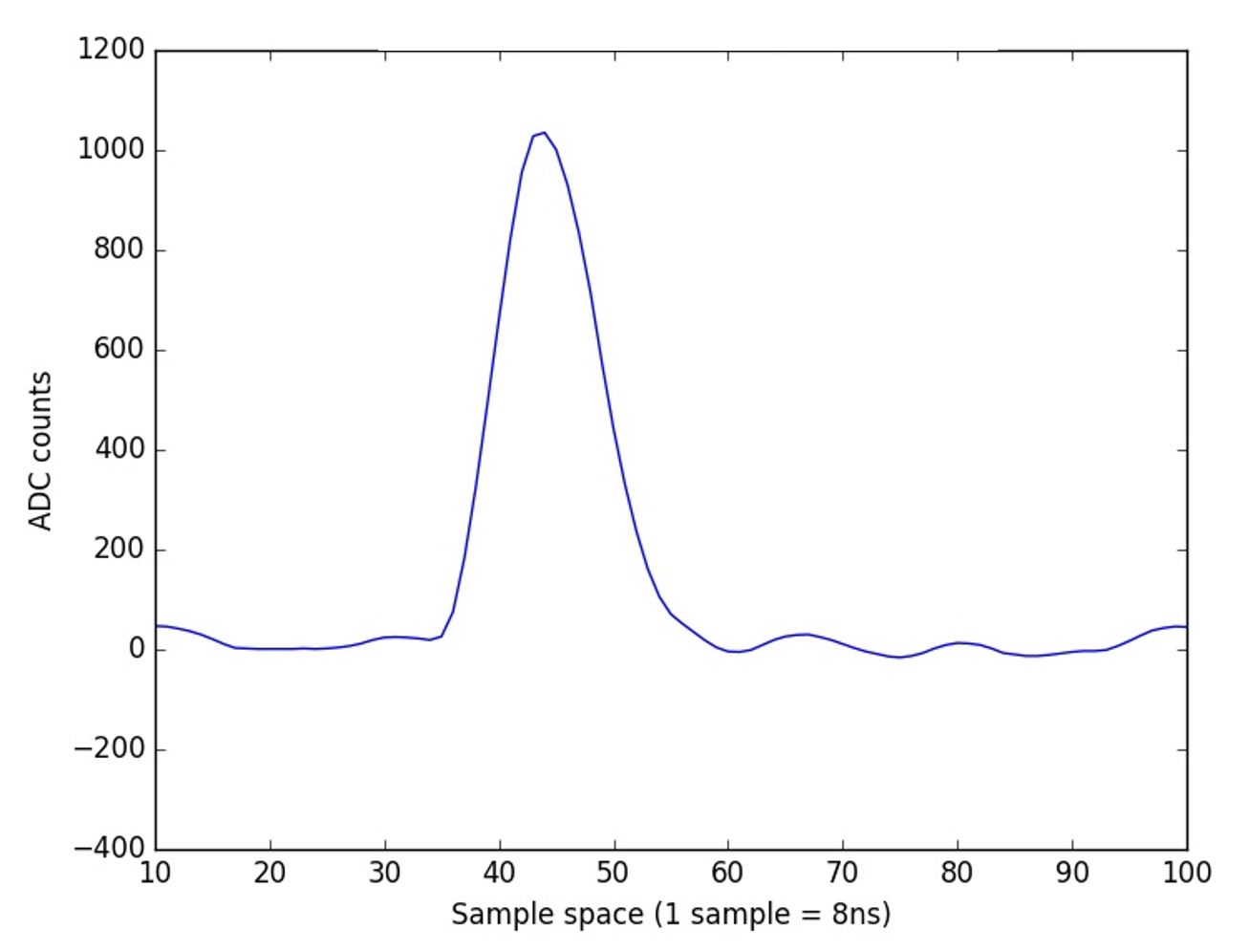}
			\caption{}
			\label{fig:1}
		\end{subfigure}%
		\begin{subfigure}{0.5\textwidth}
			\centering
			\includegraphics[width=1.2\linewidth]{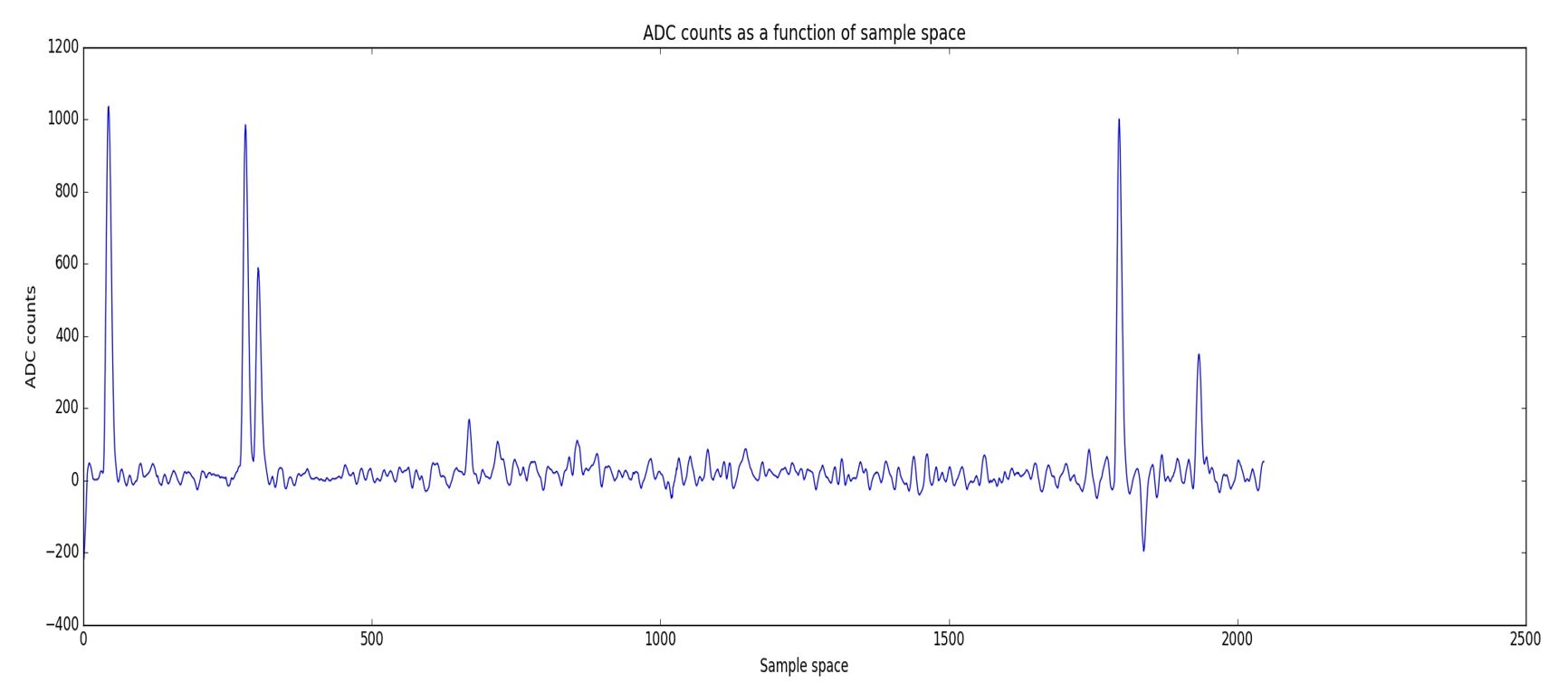}
			\caption{}
			\label{fig:2}
		\end{subfigure}
		\caption{(a)  Single channel event in sample space (b) Multiple consecutive photon traces in sample space}
		\label{fig:linearity4}
	\end{figure}

	Finally, the MCP detector was setup, biased,  and illuminated with a flux of mostly single photons.  Plots showing the temporal response in ADC counts as a function of sample space are presented in figure~\ref{fig:linearity4}.

	\newpage

	\section{Summary and future work}
	
	We developed the GRAPH ASIC, which enables the construction of a readout  system intended to couple to a large area MCP single photon imaging detector, designated for future space applications. The test board was constructed and the ASIC performance evaluated. Furthermore initial test confirm the   correct functionality as well as correct operation. In addition tests recording detector signals were carried out.  Considering the low power consumption of 47 mW per channel, an extrapolated value for a 8x8cm detector requiring 128 channels per axis *(16 ASICs) would yield a power consumption of 13 Watts FPGA excluded. In the future we look forward to testing a new kind of firmware, that would only record events, by asserting a few clock cycles only when a signal is detected via triggers, and force conversion, if an arbitrary overdue time has passed to avoid losing the signal due to leakage. In this way, we aim to optimize the usage of the memory while preserving the sampling clock. Ultimately, an improved version of HULA considering a much better SNR, could operate as an sample and hold event sampler with the ability to digitize only events. In addition, an internal precise TDC with memory could adjunct the time of arrival to the events. 
	
	The new revised version under consideration will address the noise values, as contributing sources aren't fully understood yet,  the sampling leakage issue, and reduce the amount of communication lines required. One option is certainly to buffer and serialize data of selected events. This would open the possibility to attach 20 chips on a single FPGA to equip a 100x100mm detector.

	\newpage

	\section{Acknowledgments}

	We would like to thank NASA Strategic Astrophysics Technology program (NNX12AF46G) for financial support.  We would also like to thank NALU scientific LLC for their invaluable help during the ASIC design verification process.  We would like to thank Dr. Peter Orel, Vihtori Virta, and Matt Andrew and Nalu LLC for their support.

\end{document}